%% file: main.tex
\documentclass[reprint,
 amsmath,amssymb
 prl,
]{revtex4-2}
\usepackage[caption=false]{subfig}
\usepackage{graphicx}
\usepackage{mathtools}
\usepackage{dcolumn}
\usepackage{bm}
\usepackage{amsthm}
\usepackage{amsfonts}
\usepackage{hyperref}

\begin{document}

\title{Wannier band transitions in disordered $\pi$-flux ladders}

\author{Jahan Claes}
\author{Taylor L. Hughes}
\affiliation{ Department of Physics and Institute for Condensed Matter Theory$,$\\
 University of Illinois at Urbana-Champaign$,$ Illinois 61801$,$ USA}

\begin{abstract}

Boundary obstructed topological insulators are an unusual class of higher-order topological insulators with topological characteristics determined by the so-called Wannier bands. Boundary obstructed phases can harbor hinge/corner modes, but these modes can often be destabilized by a phase transition on the boundary instead of the bulk. While there has been much work on the stability of topological insulators in the presence disorder, the topology of a disordered Wannier band, and disorder-induced Wannier transitions have not been extensively studied. In this work, we focus on the simplest example of a Wannier topological insulator: a mirror-symmetric $\pi$-flux ladder in 1D. We find that the Wannier topology is robust to disorder, and derive a real-space renormalization group procedure to understand a new type of strong disorder-induced transition between non-trivial and trivial Wannier topological phases. We also establish a connection between the Wannier topology of the ladder and the energy band topology of a related system with a physical boundary cut, something which has generally been conjectured for clean models, but has not been studied in the presence of disorder. 

\end{abstract}

\maketitle

\section{Introduction}

Topological insulators are phases of matter that cannot be deformed to a trivial atomic limit without closing the energy gap or breaking a protecting symmetry\cite{qi2011topological,hasan2010colloquium,bernevig2013topological}. Topological insulators can be protected by internal symmetries\cite{kane2005z,kane2005quantum,schnyder2008classification,kitaev2009periodic,ryu2010topological,chiu2016classification}, which results in the periodic classification table of topological insulators and superconductors\cite{schnyder2008classification,kitaev2009periodic,ryu2010topological,chiu2016classification}. They can also be protected by spatial symmetries such as reflection or rotation, leading to topological crystalline insulators (TCIs)\cite{fu2011topological,hughes2011inversion,hsieh2012topological,ando2015topological,kruthoff2017topological,po2017symmetry,bradlyn2017topological}. One important characteristic of topological insulators is the spectroscopy of their boundary states. TIs protected by internal symmetries display protected surface modes on any boundary, while TCIs can typically display protected surface modes on only boundaries that respect the spatial symmetry.

Recently, there has been interest in so-called higher-order topological insulators (HOTIs)\cite{schindler2018higher,benalcazar2017quantized,benalcazar2017electric,khalaf2019boundary}. HOTIs are crystalline insulators that display protected modes on certain corners or hinges determined by spatial symmetries. Some of these HOTIs\cite{benalcazar2017electric,schindler2018higher} are TCIs in the usual sense, in that they cannot be deformed to a trivial atomic limit without closing the energy gap or breaking the spatial symmetry. However, there also exist HOTIs that can be symmetrically deformed to a trivial atomic limit without closing the bulk gap\cite{benalcazar2017quantized,benalcazar2017electric,khalaf2019boundary}. These HOTIs have been dubbed boundary obstructed topological insulators (BOTIs) in Ref. \onlinecite{khalaf2019boundary}. A key characteristic of a BOTI is that, rather than having properties protected by an energy gap, they have properties protected by a \emph{Wannier gap} (see Sec. \ref{cleanSection}).

An important question is if topological properties are robust to disorder. In the case of TIs protected by an internal symmetry, one can generally define topological invariants that are robust to symmetry-preserving disorder, which can change only when \emph{delocalized} states appear at the Fermi level \cite{prodan2010,prodan2016non,prodan2016bulk,prodan2017computational,prodan2013non,mondragon2014topological,loring2011,schulz2016topological}. Naively, TCIs and BOTIs should not be robust to disorder, since disorder breaks the spatial symmetry. However, provided the disorder respects the spatial symmetry on average, one can still define robust topological invariants for TCIs\cite{song2015quantization,diez2015extended,mondragon2019robust,fu2012topology,fulga2014statistical} which are stable provided the system is gapped\cite{song2015quantization}. However, there has not been any study of the effects of disorder on the BOTIs.

In this article we take the first approach at characterizing BOTIs in the presence of disorder. We study a minimal model of a BOTI, what we might call a topological \emph{Wannier} insulator. Our model is a 1D $\pi$-flux ladder having Wannier band topology protected by reflection symmetries $\hat{M}_x,\hat{M}_y$. This model originated in Ref. \onlinecite{benalcazar2017quantized} where it was used as a building block for the 2D quadrupole BOTI phase. Since our model is one-dimensional, the Wannier topology does not indicate the existence of corner or hinge modes. However, the simplicity of the model allows us to study a new type of phase transition, a  disorder-induced Wannier transition, in detail.  Furthermore, since this model is a single layer of the 2D quadrupole BOTI, our work serves as a starting point for more computationally intensive studies of disordered 2D BOTIs.

We find the Wannier topology is stable to symmetry-breaking disorder provided the symmetries are obeyed on average, and there is a sharp Wannier transition at a critical value of the disorder at which the Wannier gap closes. This transition is distinct from the usual disorder-induced topological transitions, in that it occurs without delocalized states crossing the Fermi level; in fact, the system \emph{remains gapped} throughout. In addition, we find the Wannier topology is connected to the energy band topology of an artificial ``edge"  introduced by cutting the $\hat y$ bonds of the ladder to separate the two legs of the ladder. This represents the first evidence the Wannier topology gives a robust signature for the boundary topology in the presence of disorder. Finally, we introduce a real-space renormalization group approach that explains the robustness of the Wannier topology, and the connection between the Wannier gap closing and the Wannier transition. Importantly, our model can be realized in a number of experimental systems including mechanical resonator arrays\cite{serra2018}, microwave resonator arrays\cite{peterson2018}, circuit resonators\cite{imhof2018}, and cold-atoms\cite{serra2018,peterson2018,imhof2018,meier2018}. Hence we anticipate that our results will be immediately relevant for experiments in higher order topology.

\section{The clean $\pi$-flux ladder}
\label{cleanSection}
\begin{figure}
    \centering
    \includegraphics[width=\columnwidth]{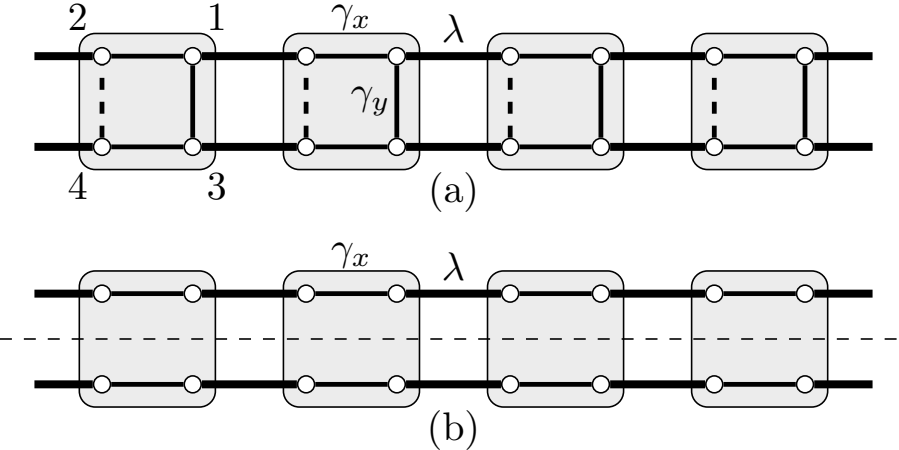}
    \caption{(a) The clean $\pi$-flux chain. Dotted lines denote negative hoppings. The pattern of negative hoppings implies a magnetic flux of $\pi$ per plaquette. (b) Cutting the $\hat y$ bonds gives two disconnected Su-Schrieffer-Heeger chains.}
    \label{model}
\end{figure}

The $\pi$-flux ladder model we use is shown in Fig. \ref{model}a. The model has four sites per unit cell, and a magnetic flux of $\pi$ through each plaquette. In terms of the intercell hopping $\lambda$ and intracell hoppings $\gamma_x,\gamma_y$, the Bloch Hamiltonian is
\begin{equation}
    \hat{H}(k)=[\gamma_x+\lambda\cos(k)]\tau_0\sigma_1+\lambda\sin(k)\tau_0\sigma_2+\gamma_y\tau_1\sigma_3,
    \label{modelHamiltonian}
\end{equation}
where $\tau_i(\sigma_i)$ are the Pauli matrices acting on the vertical(horizontal) degrees of freedom. The model has anticommuting reflection symmetries $\hat{M}_x=\tau_3\sigma_1$, $\hat{M}_y=\tau_1\sigma_0$, and the energy spectrum is
\begin{equation}
    E_k=\pm \sqrt{\gamma_x^2+2\gamma_x\lambda\cos(k)+\lambda^2+\gamma_y^2},
\end{equation}
with each energy level doubly degenerate. The spectrum is gapped provided $\gamma_y\neq 0$ or $|\gamma_x|\neq|\lambda|$.
The gapped phases of the model maintain the reflection symmetry $\hat{M}_x,$ and are classified (in 1D) by an index $\nu\in \mathbb{Z}.$ Furthermore they display a quantized electric polarization $p$ given by $2p=\nu\text{ mod }2$\cite{hughes2011inversion,alexandradinata2014wilson}. However, all gapped phases of our model are connected by some gapped path in $(\gamma_x,\gamma_y,\lambda)$, thus $\hat{H}$ can only describe a single topological phase. Indeed, the index $\nu$ of this model is zero, and $p=0$, for \emph{any} values of the parameters. We would thus classify this model as a trivial TCI.

To uncover the topological properties of this model we need a refined approach. Refs. \onlinecite{benalcazar2017quantized,benalcazar2017electric} recently introduced the idea of \emph{Wannier band topology}, a topological characterization that can change while the energy gap remains open. The idea is to divide the occupied subspace of the Hamiltonian into two separate subspaces, each of which is localized in a different part of the unit cell; these individual subspaces then have their own topological invariants. Concretely, we define the projection operator onto the occupied states, and the y-position operator: 
\begin{equation}
    \hat{P}_k  = \sum_i |u_k^i\rangle\langle u_k^i|,\quad \hat{Y}|a\rangle = \left\{\begin{smallmatrix*}[r]y_0|a\rangle & \ a=1,2\\ -y_0|a\rangle\hfil &\ a=3,4\end{smallmatrix*} \right.,
\end{equation}
where $|u_k^{1,2}\rangle$ are the two occupied eigenstates of $\hat{H}_k$ at half-filling, and $\pm y_0$ is the vertical position of the upper/lower sites. We find $\hat{P}_k\hat{Y}\hat{P}_k$ has two nonzero eigenvectors, $|w_k^\pm\rangle$, with eigenvalues $\nu_k^{\pm}$. We call $\hat{P}_k\hat{Y}\hat{P}_k$ the \emph{Wannier Hamiltonian} and $|w^\pm_k\rangle$ the \emph{Wannier bands}. Provided $g>0$, $|w_k^\pm\rangle$ is a smooth function of $k$. We can then associate a polarization to each Wannier band:
\begin{equation}
    p^{\pm}\equiv i\int \frac{dk}{2\pi} \langle w_k^\pm | \partial_k w_k^\pm\rangle .
\end{equation}
Due to $\hat{M}_x$ symmetry, $p^\pm$ is quantized to $0$ or $1/2$\cite{benalcazar2017quantized,benalcazar2017electric}, and serves as a topological invariant for the Wannier bands. Note $p^+=p^-$ due to $\hat M_y$ symmetry. In terms of $(\gamma_x,\gamma_y,\lambda)$, we have $p^\pm=1/2$ for $|\lambda|>|\gamma_x|$, and $p^\pm=0$ for $|\lambda|<|\gamma_x|$. Exactly at $|\lambda|=|\gamma_x|$, the Wannier gap closes. We see while our model is topologically trivial with respect to the usual topological invariant, it describes two topological Wannier phases separated by a Wannier gap closing.

In higher-dimensional models, the Wannier bands have been related to the edge Hamiltonian for a system with boundary\cite{fidkowski2011model,benalcazar2017quantized,benalcazar2017electric,schindler2018higher,khalaf2019boundary}. For our system, we can give a similar interpretation. Note that cutting the $\hat y$ bonds of the chain results in two isolated Su-Schrieffer-Heeger (SSH) chains\cite{su1979solitons} (Fig. \ref{model}b). The topology of the upper/lower Wannier band is identical to the topology of the upper/lower SSH chain that results when $\gamma_y$ is set to zero. This follows because tuning $\gamma_y$ to zero cannot close the Wannier gap or the energy gap, thus the Wannier polarizations cannot change during the transition. When $\gamma_y=0$, the upper/lower Wannier bands are precisely the ground states of the upper/lower SSH chains. Therefore, the polarization of the Wannier bands are identical to the polarization of the corresponding SSH chains. 

\section{Disorder-induced transitions in the $\pi$-flux ladder}

To study the effect of disorder on Wannier band topology, we randomly perturb each link in our Hamiltonian,
\begin{equation}
    \gamma_{x,y}^n = \gamma_{x,y}+W_\gamma\omega_{x,y}^n, \qquad
    \lambda^n = \lambda+W_\lambda\omega^n,
\end{equation}
where the $\omega^n_{(x,y)}\in [0,1]$ are uniformly distributed random variables, and $(W_\gamma,W_\lambda)$ parameterize the disorder strength. Note that we choose the link disorder to be positive rather than symmetric about zero; this ensures the disordered model still has $\pi$-flux through each plaquette. While maintaining $\pi$-flux is not essential for our conclusions, this choice separates transitions where the Wannier gap closes from the ones where the energy gap closes, thus allowing us to isolate the Wannier transition. While our disorder breaks $\hat{M}_x$ and $\hat{M}_y$ symmetries, there is still a sense in which it approximately respects these symmetries. We say a symmetry $\hat{S}$ is respected \emph{on average} if a disordered Hamiltonian $H$ occurs with equal probability as $\hat{S}^\dagger \hat{H} \hat{S}$\cite{fulga2014statistical}. We see that our disordered Hamiltonians indeed respect these symmetries on average.

In the presence of disorder we can define Wannier topology by introducing the analogous operators
\begin{equation}
    \hat{P}  = \sum_n |\psi_n\rangle\langle\psi_n|,\quad     \hat{Y}|x,a\rangle = \left\{\begin{smallmatrix*}[r]y_0|x,a\rangle & \ a=1,2\\ -y_0|x,a\rangle\hfil &\ a=3,4\end{smallmatrix*} \right.,
\end{equation}
where $|\psi_n\rangle$ are the occupied states of $H$. The upper/lower Wannier bands are the eigenstates $|w_n^\pm\rangle$ of the Wannier Hamiltonian $\hat P \hat Y\hat P$ with corresponding Wannier values $\nu_n^\pm$. We can define the polarization of the upper/lower Wannier band $p^\pm$ either by the method of Refs. \onlinecite{schulz2013orbital,song2015quantization}, or the method of Ref. \onlinecite{resta1998quantum}. For simplicity, we follow Ref. \onlinecite{resta1998quantum} and define
\begin{equation}
    z^\pm \equiv \det(\hat{P}^\pm e^{i\alpha \hat{X}} \hat{P}^\pm),\qquad p^\pm = \frac{1}{2\pi}\text{arg} (z^\pm),
\label{WannierPolarizationDefn}
\end{equation}
where $\hat{P}^\pm$ is the projector onto the upper/lower Wannier band, $\alpha\equiv 2\pi/L_x$, and $L_x$ is the length of the system in the $x$-direction. We can also define the localization length of states in the Wannier bands\cite{resta1999electron}
\begin{equation}
\Lambda^\pm = \frac{1}{2\pi}\sqrt{-L_x\log|z^\pm|^2}.
\label{LocalizationLengthDefn}
\end{equation}

It has been proven that as long as a Hamiltonian is local, gapped, and respects $\hat{M}_x$ and translation symmetry on average, the $x$-polarization is self-averaging and quantized to $0$ or $1/2$\cite{song2015quantization}.  This result, applied to the Wannier Hamiltonian $\hat{P}\hat{Y}\hat{P}$, implies that the Wannier polarizations are self-averaging and quantized to $0$ or $1/2$ in the presence of a Wannier gap. To see this, we note that the Wannier Hamiltonian  $\hat{P}\hat{Y}\hat{P}$ is local provided our original Hamiltonian is gapped, and that a projector $\hat{P}$ is equally likely to occur as its symmetry-related counterpart $\hat{M}_x\hat{P}\hat{M}_x$. Finally, since $\hat{P}\hat{Y}\hat{P}$ respects $\hat{M}_y$ symmetry on average, we can also conclude $p^+=p^-$ even with disorder.

\begin{figure}
    \centering
    \includegraphics[width=\columnwidth]{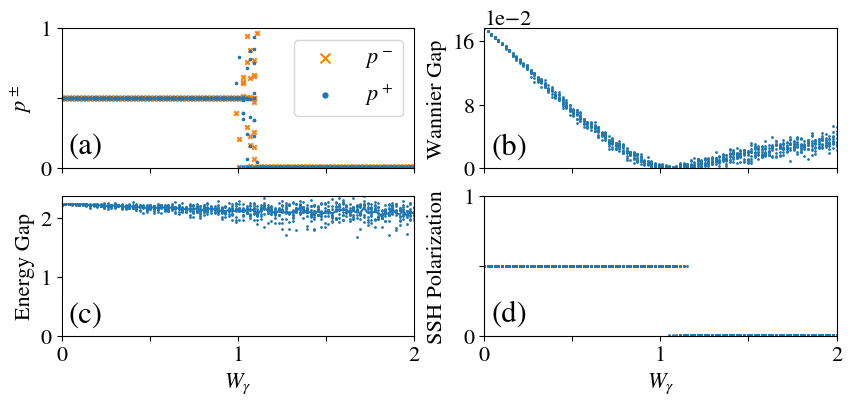}
    \caption{(a) $p^\pm$ as a function of $W_\gamma$. (b) The Wannier gap as a function of $W_\gamma$. We find that $p^\pm$ changes exactly when the Wannier gap closes. (c) The bulk energy gap, which we find remains open, allowing us to study only the Wannier topology. (d) The polarization of one of the SSH chains obtained by removing the vertical links (rungs), as in Fig. \ref{model}b. We see the SSH polarization and the Wannier polarization agree. Each panel has $L_x=300$ and 20 disorder realizations at each $W_\gamma$.}
    \label{disorderResults}
\end{figure}

An example of this characteristic behavior is shown in Fig. \ref{disorderResults}. Here we set $(\gamma_x,\gamma_y,\lambda)=(.5,1,1)$, and tune only $W_\gamma$. In Figs. \ref{disorderResults}(a,b), we plot $p^\pm$ as a function of $W_\gamma$. In Fig. \ref{disorderResults}b, we plot the Wannier gap as a function of $W_\gamma$. We see that, as predicted, $p^\pm$ are quantized to $0$ or $1/2$, and they change at only the point where the Wannier gap closes. For comparison, we show the energy gap in Fig. \ref{disorderResults}c, which remains open during this process.

\begin{figure}
    \centering
    \includegraphics[width=\columnwidth]{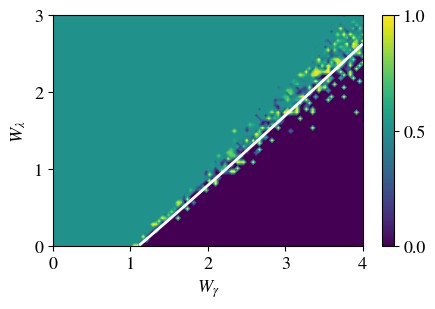}
    \caption{$p^+$ as a function of $(W_\gamma,W_\lambda)$. The color denotes the direct calculation of $p^+$ for $L_x=300$, while the overlaid line is the analytically predicted topological transition of the corresponding SSH chain\cite{mondragon2014topological}. We see that the Wannier transition occurs exactly when the SSH chain has a topological transition.}
    \label{phaseDiagram}
\end{figure}

Intriguingly, we also find that the Wannier band polarization and the polarization of the isolated upper/lower SSH chains agree even with disorder, as shown in Fig. \ref{disorderResults}d where we plot the polarization of the upper chain as a function of $W_\gamma$. Here, we are comparing the polarization of the Wannier bands with the polarization of our model at the same parameter values with the vertical bonds (rungs) turned off. For our model, with only bond disorder, we can use the results of \cite{mondragon2014topological} to analytically predict the phase diagram of the  decoupled upper/lower SSH chains, and compare it to the calculated nested Wannier polarizations. The result is shown in Fig. \ref{phaseDiagram}. Here, the line denotes the exact phase boundary for a single SSH chain, while the color indicates the nested polarization for the $\pi$-flux ladder. We see that the transition of the upper SSH chain exactly agrees with the Wannier transition, and thus we can analytically predict the location of the Wannier transition when the chains are coupled. We find this result actually holds for more general Hamiltonians, such as those including random on-site disorder or next-nearest-neighbor hoppings that preserve the symmetries on average, though we cannot analytically determine the phase diagram of the decoupled chains in these cases. From this we can conclude that, even in the presence of disorder, the Wannier Hamiltonian describes the topological properties of the physical system with a spatial cut.

\section{Renormalization Group picture of the transition}
\label{RGSection}
\begin{figure}
    \centering
    \includegraphics[width=\columnwidth]{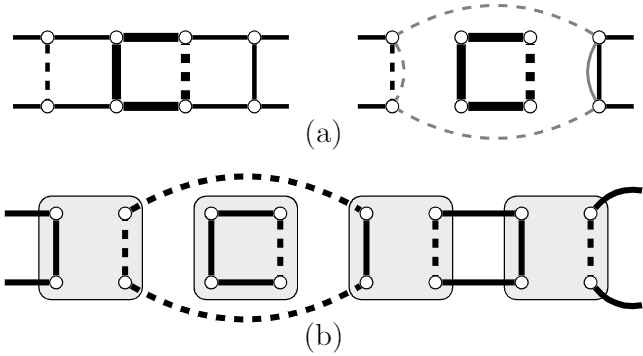}
    \caption{(a) The RG procedure. We take the strongest plaquette of the ladder, and project into its local ground state. This generates new effective couplings between previously unconnected sites in a way that preserves the structure of the ladder. (b) At the end of the RG procedure, the system is approximated by disconnected $\pi$-flux plaquettes.}
    \label{renormalization}
\end{figure}

We can gain additional insight into the nature of the Wannier transition through a real-space renormalization group (RG) procedure. Our RG procedure is illustrated in Fig. \ref{renormalization}a, and is based on the real-space RG introduced in Ref. \onlinecite{fisher1994random} to study random spin chains. Indeed, a similar method has been applied to disordered 1D topological insulators (SSH chains with chiral symmetry) in Ref. \onlinecite{mondragon2014topological}. To construct the RG procedure we first define the strength $s$ of a single plaquette, which we consider to be the ratio of its energy gap to the largest $x$-bond connected to it. The RG step is to choose the strongest plaquette, and project onto its local ground state. To lowest order, the projection generates the new (weaker) effective hoppings shown in Fig. \ref{renormalization}a. Because of the sublattice symmetry, the RG does not generate diagonal couplings. In addition, the signs of the effective couplings preserve the $\pi$-flux around the remaining plaquettes. We note that this is true for all relevant $\pi$-flux gauge choices, not just the gauge shown (see Supplement
\footnote{See Supplemental Material at [URL will be inserted by publisher] for a detailed description of the RG procedure and proof that the RG preserves the structure of the $\pi$-flux ladder}
for details). Thus, after one RG step we are left with another $\pi$-flux ladder with $4$ fewer sites. Iterating, we eventually reach a point where our system consists entirely of disconnected plaquettes (Fig. \ref{renormalization}b). Each step of the RG is valid provided the plaquette strength is large; we expect that at strong disorder, the distribution of hoppings is broad enough that at every step there will be a plaquette with $s>>1$.

\begin{figure}
    \centering
    \includegraphics[width=\columnwidth]{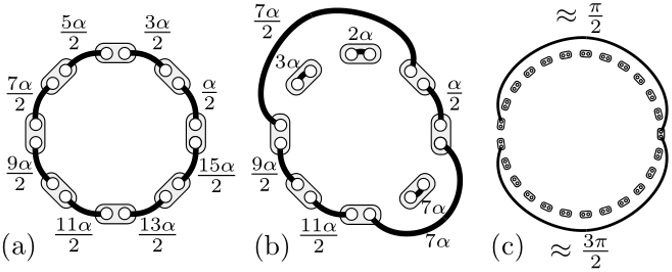}
    \caption{(a) A schematic illustration for the plaquettes after RG for a clean $8$-site system with $p^\pm=.5$. Each circle represents two sites and each bond represents a plaquette. The number near each bond is the plaquette's contribution to $\text{arg}(z^\pm)$. (b) A local rearrangement of plaquettes changes the amount each plaquette contributes to $\text{arg}(z^\pm)$, but cannot change the total $\text{arg}(z^\pm)$. (c) A bond of length $\sim L/2$ can be locally rearranged to change $\text{arg}(z^\pm)$ by $\pi$. This is the only way for $p^\pm$ to change.}
    \label{computingP}
\end{figure}

Once our system is approximated by disconnected plaquettes, we can compute the Wannier polarization $p^\pm$ via Eq. \ref{WannierPolarizationDefn}. At half filling, direct calculation shows that each plaquette has a Wannier gap unless the horizontal bonds are both zero, and the Wannier functions $|w_n^\pm\rangle$ have equal weight on the left and right sides of the plaquette. It follows that a plaquette that stretches between unit cells $a$ and $b$ contributes a phase of $e^{\frac{i\alpha(a+b)}{2}}$ to $z^\pm$. In Fig. \ref{computingP}a, we schematically illustrate the plaquettes for a clean system in the topological phase and the contribution of each plaquette to $\arg(z^\pm)$. Introducing disorder locally rearranges the plaquettes, as in Fig. \ref{computingP}b, but local rearrangements do not change $\arg(z^\pm)$. The only way for a rearrangement to change $p^\pm$ is for a plaquette to grow to length $L_x/2$, which can change the sign of $z^\pm$ and thus change $p^\pm$ by $1/2$, as in Fig. \ref{computingP}c. The RG alsdemonstrates the critical point is characterized by a diverging Wannier localization length $\Lambda^\pm$. From Eq. \ref{LocalizationLengthDefn}, if our plaquettes span sites $a_j$ and $b_j$, the localization length is
\begin{equation}
    \Lambda^\pm = \frac{1}{2\pi}\sqrt{L_x\sum_j \log[\csc^2(\pi\frac{a_j-b_j}{L_x})]}
\end{equation}
which diverges when one of the $(a_j-b_j)=L_x/2.$ 

Heuristically, because of the nature of the RG, plaquettes spanning a larger width have weaker bonds in the $x$-direction. In the thermodynamic limit, a plaquette of width $~L_x/2$ will have vanishing hopping in the $x$-direction, and the plaquette will effectively have only vertical bonds. This configuration makes the Wannier gap vanish (electrons are exactly halfway between the upper and lower chains). The RG establishes a clear connection between the Wannier gap closing and the topological phase transition; the phase transition happens when long-range plaquettes form, which have vanishing Wannier gap.

\section{Discussion}

The $\pi$-flux ladder possesses a robust Wannier topological invariant, $p^\pm$, protected by the mirror symmetries $\hat{M}_x,\hat{M}_y$. With disorder, the Wannier topological invariant remains quantized provided the energy and Wannier gaps remain open. The physical interpretation of the Wannier topological invariant in 1D is subtle, since, for example, there are no corners to display robust mid-gap modes or fractional charges. However, we have seen that introducing an artificial ``edge" into the system connects the Wannier topology and the edge topology for both clean and disordered systems. This conclusion holds numerically for general local symmetry-preserving disorder, and not just the link disorder presented here. We conjecture that this is because tuning the $y$-bonds to zero cannot close the energy or Wannier gap provided the Hamiltonian obeys $\hat{M}_x$ and $\hat{M}_y$ symmetries on average. This work thus provides the first evidence of a robust connection between Wannier and edge topology in the presence of disorder. We anticipate that our results can be immediately tested in metamaterial or cold atom experiments.\cite{serra2018,peterson2018,imhof2018,meier2018}

Because of the simplified nature of our model, we can predict the behavior of the ``edge" from first principles and confirm the connection between Wannier and edge topology. We can also understand the Wannier transition through a real-space RG, which offers a qualitative explanation of the local stability to disorder and the connection between the Wannier transition, Wannier delocalization, and Wannier gap closing.

While these results form the first example of the stability of Wannier topology to disorder, ultimately it will be useful to generalize these results to higher-dimensional models, where nontrivial Wannier topology implies protected corner or hinge modes. It will be especially interesting to see if the connection between Wannier topology and edge topology remains for higher-dimensional models.

\acknowledgements
TLH thanks W. A. Benalcazar, E. Khalaf, and R. Queiroz for useful discussions on BOTIs. TLH thanks the US National Science Foundation for support under the award DMR 1351895-CAR, and the NSF MRSEC program under NSF Award Number DMR-1720633 (SuperSEED).

\bibliography{thebibliography.bib}

\input{Supplement}

\end{document}

%% file: Supplement.tex
\onecolumngrid

\renewcommand{\thesection}{}
\setcounter{table}{0}
\renewcommand{\thetable}{S\arabic{table}}
\setcounter{figure}{0}
\renewcommand{\thefigure}{S\arabic{figure}}
\setcounter{equation}{0}
\renewcommand{\theequation}{S\arabic{equation}}

\section*{Supplemental Material}
\label{RGAppendix}
\def \scale {.36}

\begin{figure}
    \includegraphics[scale=\scale]{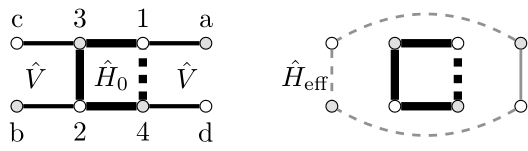}
    \caption{The setup for computing the effective Hamiltonian. We restrict attention to only hopping elements that affect the central plaquette. Our Hamiltonian has a sublattice symmetry with respect to the two sublattices A/B indicated by the shaded/unshaded circles. We treat the hoppings within the central plaquette as our unperturbed Hamiltonian $\hat{H}_0$, and the hoppings from the neighboring sites (a)-(d) as a perturbation $\hat{V}$. The resulting effective Hamiltonian is computed to second order in $\hat V$ and involves only sublattice-symmetric hoppings between sites (a)-(d).}
    \label{EffectiveHam}
\end{figure}

In this Supplement, we give the details of the local projection involved in the RG, and prove that the resulting effective Hamiltonian is again a $\pi$-flux ladder. We follow the procedure outlined in Ref. \onlinecite{fisher1994random} for finding the effective couplings after projecting into a local ground state. We consider the Hamiltonian for the eight sites shown in Fig. \ref{EffectiveHam}, neglecting the rest of the ladder. We call the hoppings on the central plaquette $\hat{H}_0$, and the hoppings between the central plaquette and sites (a)-(d) $\hat V$. Note that our total Hamiltonian has a sublattice symmetry, in that we can divide our lattice into two sublattices, A and B, such that all hopping takes place between the two sublattices (in the figure, the sublattice A and B are denoted by shaded and unshaded circles, respectively).

To find the effective Hamiltonian, we treat $\hat{V}$ as a perturbation and work to second order in $\hat{V}$; this is valid provided the energy gap is much larger than the hoppings. Then the general formula for the effective Hamiltonian is given by\cite{mila2011strong}
\begin{equation}
    \hat{H}_{\text{eff}}=E_g+\hat{P}\hat{V}\hat{P}+\hat{P}\hat{V}(E_g-\hat{Q}\hat{H}_0\hat{Q})^{-1}\hat{Q}\hat{V}\hat{P}
\end{equation}
where $E_g$ is the ground state energy of $\hat{H}_0$, $\hat{P}$ is the projector onto the degenerate ground state subspace of $\hat{H}_0$, and $\hat{Q}=1-\hat{P}$. In our case, we can make a few simplifications. First, since we are only interested in the ground state wavefunctions and not their energies, we can neglect the constant $E_g$. Second, our particular $\hat{V}$ satisfies $\hat{P}\hat{V}\hat{P}=0$, since every state in the range of $\hat{P}$ has two electrons in the central plaquette, and $\hat{V}$ always changes the number of electrons in the central plaquette by one. For the same reason, we have $\hat{Q}\hat{V}\hat{P}=\hat{V}\hat{P}$. Thus in total, our simplified equation becomes
\begin{equation}
    \hat{H}_{\text{eff}}=\hat{P}\hat{V}(E_g-\hat{Q}\hat{H}_0\hat{Q})^{-1}\hat{V}\hat{P}.
    \label{effectiveHamEq}
\end{equation}

We need to prove three things about $H_\text{eff}$:
\begin{enumerate}
    \item $H_\text{eff}$ is a non-interacting Hamiltonian, i.e. it contains only hopping and on-site terms.
    \item $H_\text{eff}$ has the same sublattice symmetry as the original Hamiltonian, so that there are no on-site terms or hoppings from (a) to (c) or from (b) to (d).
    \item $H_\text{eff}$ has $\pi$-flux through the plaquette spanned by (a)-(d) and combines with the rest of the ladder to create another $\pi$-flux ladder.
\end{enumerate}

To prove these results, we'll derive an explicit formula for $\hat{H}_{\text{eff}}$ in terms of $\hat{V}$ and $\hat{H}_0$ that makes no reference to $\hat{P}$ or $\hat{Q}$. We start by rewriting Eq. \ref{effectiveHamEq} in second-quantized notation. We'll define operators $\hat{c}_\alpha^{(\dagger)}, \alpha\in \{a,b,c,d\}$ to be the annihilation (creation) operators for the sites outside the central plaquette, and $\gamma_i^{(\dagger)}, i\in \{1,2,3,4\}$ to be the annihilation (creation) operators for the $i$th energy mode on the central plaquette with energy $E_i$. We can write a second quantized basis of states as $|abcdn_1n_2n_3n_4\rangle$ with $a$-$d$ being the occupation numbers of sites (a)-(d), and $n_i$ being the occupation number of the $i$th energy mode on the central plaquette. In terms of these operators, we can write $\hat{V}$ as
\begin{equation}
    \hat{V}=\sum_{\alpha, i} \left[\hat{c}_\alpha^\dagger V_{\alpha i} \hat{\gamma}_i +\hat{\gamma}_i^\dagger V_{i\alpha}^\dagger \hat{c}_\alpha\right]
    \label{definitionV}
\end{equation}
where $V_{\alpha i}$ is some matrix. In this notation, $[\hat{c}_\alpha^{(\dagger)},\hat{P}]=0$, and the effect of $\hat{P}$ is to project to the orthonomal basis $\{|abcd1100\rangle | a,b,c,d=0,1\}$.

We can then write the action of $H_{\text{eff}}$ on any state in the range of $\hat{P}$ as (we ignore terms that involve $\hat{\gamma}_i\hat{\gamma}_j$ or $\hat{\gamma}_i^\dagger\hat{\gamma}_j^\dagger$ since such terms take us to states orthogonal to $\hat{P}$)
\begin{equation}
    \hat{H}_{\text{eff}}=\sum_{i,j,\alpha,\beta}\left[\hat{c}_\alpha^\dagger V_{\alpha i} \hat{P}\hat{\gamma}_i(E_g-\hat{Q}\hat{H}_0\hat{Q})^{-1}\hat{\gamma}_j^\dagger V_{j\beta}^\dagger \hat{c}_\beta+\hat{c}_\beta V_{j\beta}^\dagger \hat{P}\hat{\gamma}_j^\dagger (E_g-\hat{Q}\hat{H}_0\hat{Q})^{-1}\hat{\gamma}_i V_{\alpha i}\hat{c}_\alpha^\dagger \right].
\end{equation}
Now, when acting on a state of the form $|abcd1100\rangle$, the operator $\hat{P}\hat{\gamma}_i(E_g-\hat{Q}\hat{H}_0\hat{Q})^{-1}\hat{\gamma}_j^\dagger$ has a simple action:
\begin{equation}
    \hat{P}\hat{\gamma}_i(E_g-\hat{Q}\hat{H}_0\hat{Q})^{-1}\hat{\gamma}_j^\dagger =\left(\begin{smallmatrix}0&0&0&0\\0&0&0&0\\0&0&-\frac{1}{E_3}&0\\0&0&0&-\frac{1}{E_4}\\\end{smallmatrix}\right)_{ij}.
\end{equation}
Similarly,
\begin{equation}
    \hat{P}\hat{\gamma}_j^\dagger(E_g-\hat{Q}\hat{H}_0\hat{Q})^{-1}\hat{\gamma}_i =\left(\begin{smallmatrix}\frac{1}{E_1}&0&0&0\\0&\frac{1}{E_2}&0&0\\0&0&0&0\\0&0&0&0\\\end{smallmatrix}\right)_{ij}.
\end{equation}
Thus in total, we can write
\begin{align}
    \hat{H}_{\text{eff}}=&\sum_{i,j,\alpha,\beta}\left[\hat{c}_\alpha^\dagger V_{\alpha i} \left(\begin{smallmatrix}0&0&0&0\\0&0&0&0\\0&0&-\frac{1}{E_3}&0\\0&0&0&-\frac{1}{E_4}\\\end{smallmatrix}\right)_{ij}V_{j\beta}^\dagger \hat{c}_\beta+\hat{c}_\beta V_{j\beta}^\dagger  \left(\begin{smallmatrix}\frac{1}{E_1}&0&0&0\\0&\frac{1}{E_2}&0&0\\0&0&0&0\\0&0&0&0\\\end{smallmatrix}\right)_{ij}V_{\alpha i}\hat{c}_\alpha^\dagger \right]\\
    =&\sum_{i,j,\alpha,\beta}\left[\hat{c}_\alpha^\dagger V_{\alpha i} \left(\begin{smallmatrix}0&0&0&0\\0&0&0&0\\0&0&-\frac{1}{E_3}&0\\0&0&0&-\frac{1}{E_4}\\\end{smallmatrix}\right)_{ij}V_{j\beta}^\dagger \hat{c}_\beta+ \hat{c}_\alpha^\dagger V_{\alpha i}\left(\begin{smallmatrix}-\frac{1}{E_1}&0&0&0\\0&-\frac{1}{E_2}&0&0\\0&0&0&0\\0&0&0&0\\\end{smallmatrix}\right)_{ij}V_{j\beta}^\dagger\hat{c}_\beta +\delta_{\alpha\beta}V_{\alpha i}\left(\begin{smallmatrix}\frac{1}{E_1}&0&0&0\\0&\frac{1}{E_2}&0&0\\0&0&0&0\\0&0&0&0\\\end{smallmatrix}\right)_{ij}V_{j\beta}^\dagger\right]\\   
    =&-\sum_{i,\alpha,\beta}\left[\hat{c}_\alpha^\dagger V_{\alpha i} \frac{1}{E_i}V_{i\beta}^\dagger \hat{c}_\beta\right]
    \label{HeffManyBody}
\end{align}
where in the last line, we have dropped an irrelevant constant term. We thus clearly see that $H_{\text{eff}}$ is non-interacting.

Since we now know all our operators $\hat{H}_0$, $\hat{V}$, and $\hat{H}_{\text{eff}}$ are non-interacting, it is more convenient going forward to work with one-body operators rather than the many-body, second-quantized operators. We'll denote one-body operators using the same symbols as our many-body operators, with the understanding that in what follows we are working only with the one-body operators. We divide our full Hilbert space into two subspaces, $\mathcal{H} = \mathcal{H}_s\oplus\mathcal{H}_p$, where $\mathcal{H}_s$ is the Hilbert space of sites (a)-(d), and $\mathcal{H}_p$ is the Hilbert space of the central plaquette. Then we define $\hat{V}:\mathcal{H}_s\rightarrow\mathcal{H}_p$ to be all hoppings from sites (a)-(d) to the central plaquette, so that $\hat{V}^\dagger:\mathcal{H}_p\rightarrow\mathcal{H}_s$ consists of all hoppings in the reverse direction. We define $\hat{H}_0:\mathcal{H}_p\rightarrow\mathcal{H}_p$ to be all hoppings within the central plaquette. Note that by restricting $\hat{H}_0$ to $\mathcal{H}_p$ rather than the full Hilbert space, we've ensured $\hat{H}_0$ is invertible. Finally, we define the one-body $\hat{H}_{\text{eff}}:\mathcal{H}_s\rightarrow\mathcal{H}_s$ to be the restriction of the effective Hamiltonian to $\mathcal{H}_s$. In terms of these one-body operators, Eq. \ref{HeffManyBody} becomes
\begin{equation}
    \hat{H}_{\text{eff}}=-\hat{V}^\dagger\hat{H}_0^{-1}\hat{V}-\hat{V}\hat{H}_0^{-1}\hat{V}^\dagger.
    \label{HeffOneBody}
\end{equation}

To prove that $\hat{H}_{\text{eff}}$ has sublattice symmetry, we note that $\hat{V}$, $\hat{V}^\dagger$, and $\hat{H}_0^{-1}$ all have sublattice symmetry, and the product of three sublattice-symmetric operators is also sublattice-symmetric. Thus from Eq. \ref{HeffOneBody}, we immediately see $\hat{H}_{\text{eff}}$ has sublatttice symmetry.

Finally, we need to prove that the RG preserves the $\pi$-flux structure. We first note that $\hat{H}_0$ has the form
\begin{equation}
    \hat{H}_0=\left(\begin{smallmatrix}0&0&+&-\\0&0&+&+\\+&-&0&0\\-&+&0&0\end{smallmatrix}\right)
\end{equation}
Inspection shows that $\hat{H}_0^{-1}$ also has the same sign structure:
\begin{equation}
    \hat{H}_0^{-1}=\left(\begin{smallmatrix}0&0&+&-\\0&0&+&+\\+&+&0&0\\-&+&0&0\end{smallmatrix}\right)
\end{equation}
Then from Eq. \ref{HeffOneBody} and the fact that all the hoppings in $\hat{V}$ are positive we immediately see the signs of the hoppings in $\hat{H}_\text{eff}$ are as shown in Fig. \ref{EffectiveHam}.

\begin{figure}
    \subfloat[]{\includegraphics[scale=\scale]{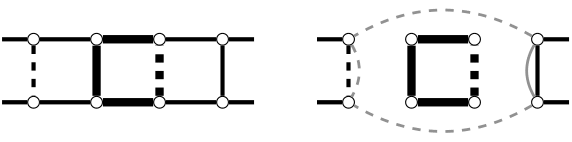}\label{sign1}}
    
    \subfloat[]{\includegraphics[scale=\scale]{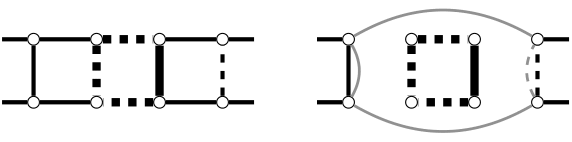}\label{sign2}}
    
    \subfloat[]{\includegraphics[scale=\scale]{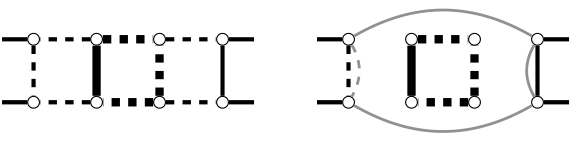}\label{sign3}}
    \caption{The RG generates $\pi$-flux ladders with all sign structures compatible with $\pi$-flux that are both real and $\hat{M}_y$ symmetric. For each of these sign structures, the RG results in induced couplings that reinforce existing couplings and result in a new $\pi$-flux ladder.}
    \label{allSignStructures}
\end{figure}

This sign structure ensures that the new induced hoppings reinforce the existing couplings to maintain a $\pi$-flux structure in the rest of the ladder, as shown in Fig. \ref{sign1}. As the RG progresses, the central plaquette and neighboring bonds can have other sign structures, two examples of which are shown in Figs. \ref{allSignStructures}b,c. The sign structures are always real and symmetric with respect to $\hat{M}_y$. A similar analysis shows that in all cases, the resulting induced hoppings always reinforce the existing couplings and maintain the overall $\pi$-flux structure. This can be seen just by inspecting the figures, using Eq. \ref{HeffOneBody} and the fact that $\hat{H}_0^{-1}$ always has the same sign structure as $\hat{H}_0$.

%% file: main.bbl
\begin{thebibliography}{46}%
\makeatletter
\providecommand \@ifxundefined [1]{%
 \@ifx{#1\undefined}
}%
\providecommand \@ifnum [1]{%
 \ifnum #1\expandafter \@firstoftwo
 \else \expandafter \@secondoftwo
 \fi
}%
\providecommand \@ifx [1]{%
 \ifx #1\expandafter \@firstoftwo
 \else \expandafter \@secondoftwo
 \fi
}%
\providecommand \natexlab [1]{#1}%
\providecommand \enquote  [1]{``#1''}%
\providecommand \bibnamefont  [1]{#1}%
\providecommand \bibfnamefont [1]{#1}%
\providecommand \citenamefont [1]{#1}%
\providecommand \href@noop [0]{\@secondoftwo}%
\providecommand \href [0]{\begingroup \@sanitize@url \@href}%
\providecommand \@href[1]{\@@startlink{#1}\@@href}%
\providecommand \@@href[1]{\endgroup#1\@@endlink}%
\providecommand \@sanitize@url [0]{\catcode `\\12\catcode `\$12\catcode
  `\&12\catcode `\#12\catcode `\^12\catcode `\_12\catcode `\%12\relax}%
\providecommand \@@startlink[1]{}%
\providecommand \@@endlink[0]{}%
\providecommand \url  [0]{\begingroup\@sanitize@url \@url }%
\providecommand \@url [1]{\endgroup\@href {#1}{\urlprefix }}%
\providecommand \urlprefix  [0]{URL }%
\providecommand \Eprint [0]{\href }%
\providecommand \doibase [0]{https://doi.org/}%
\providecommand \selectlanguage [0]{\@gobble}%
\providecommand \bibinfo  [0]{\@secondoftwo}%
\providecommand \bibfield  [0]{\@secondoftwo}%
\providecommand \translation [1]{[#1]}%
\providecommand \BibitemOpen [0]{}%
\providecommand \bibitemStop [0]{}%
\providecommand \bibitemNoStop [0]{.\EOS\space}%
\providecommand \EOS [0]{\spacefactor3000\relax}%
\providecommand \BibitemShut  [1]{\csname bibitem#1\endcsname}%
\let\auto@bib@innerbib\@empty
\bibitem [{\citenamefont {Qi}\ and\ \citenamefont
  {Zhang}(2011)}]{qi2011topological}%
  \BibitemOpen
  \bibfield  {author} {\bibinfo {author} {\bibfnamefont {X.-L.}\ \bibnamefont
  {Qi}}\ and\ \bibinfo {author} {\bibfnamefont {S.-C.}\ \bibnamefont {Zhang}},\
  }\bibfield  {title} {\bibinfo {title} {Topological insulators and
  superconductors},\ }\href@noop {} {\bibfield  {journal} {\bibinfo  {journal}
  {Rev. Mod. Phys.}\ }\textbf {\bibinfo {volume} {83}},\ \bibinfo {pages}
  {1057} (\bibinfo {year} {2011})}\BibitemShut {NoStop}%
\bibitem [{\citenamefont {Hasan}\ and\ \citenamefont
  {Kane}(2010)}]{hasan2010colloquium}%
  \BibitemOpen
  \bibfield  {author} {\bibinfo {author} {\bibfnamefont {M.~Z.}\ \bibnamefont
  {Hasan}}\ and\ \bibinfo {author} {\bibfnamefont {C.~L.}\ \bibnamefont
  {Kane}},\ }\bibfield  {title} {\bibinfo {title} {Colloquium: topological
  insulators},\ }\href@noop {} {\bibfield  {journal} {\bibinfo  {journal} {Rev.
  Mod. Phys.}\ }\textbf {\bibinfo {volume} {82}},\ \bibinfo {pages} {3045}
  (\bibinfo {year} {2010})}\BibitemShut {NoStop}%
\bibitem [{\citenamefont {Bernevig}\ and\ \citenamefont
  {Hughes}(2013)}]{bernevig2013topological}%
  \BibitemOpen
  \bibfield  {author} {\bibinfo {author} {\bibfnamefont {B.~A.}\ \bibnamefont
  {Bernevig}}\ and\ \bibinfo {author} {\bibfnamefont {T.~L.}\ \bibnamefont
  {Hughes}},\ }\href@noop {} {\emph {\bibinfo {title} {Topological insulators
  and topological superconductors}}}\ (\bibinfo  {publisher} {Princeton
  university press},\ \bibinfo {year} {2013})\BibitemShut {NoStop}%
\bibitem [{\citenamefont {Kane}\ and\ \citenamefont
  {Mele}(2005{\natexlab{a}})}]{kane2005z}%
  \BibitemOpen
  \bibfield  {author} {\bibinfo {author} {\bibfnamefont {C.~L.}\ \bibnamefont
  {Kane}}\ and\ \bibinfo {author} {\bibfnamefont {E.~J.}\ \bibnamefont
  {Mele}},\ }\bibfield  {title} {\bibinfo {title} {${Z}_2$ topological order
  and the quantum spin hall effect},\ }\href@noop {} {\bibfield  {journal}
  {\bibinfo  {journal} {Phys. Rev. Lett.}\ }\textbf {\bibinfo {volume} {95}},\
  \bibinfo {pages} {146802} (\bibinfo {year} {2005}{\natexlab{a}})}\BibitemShut
  {NoStop}%
\bibitem [{\citenamefont {Kane}\ and\ \citenamefont
  {Mele}(2005{\natexlab{b}})}]{kane2005quantum}%
  \BibitemOpen
  \bibfield  {author} {\bibinfo {author} {\bibfnamefont {C.~L.}\ \bibnamefont
  {Kane}}\ and\ \bibinfo {author} {\bibfnamefont {E.~J.}\ \bibnamefont
  {Mele}},\ }\bibfield  {title} {\bibinfo {title} {Quantum spin hall effect in
  graphene},\ }\href@noop {} {\bibfield  {journal} {\bibinfo  {journal} {Phys.
  Rev. Lett.}\ }\textbf {\bibinfo {volume} {95}},\ \bibinfo {pages} {226801}
  (\bibinfo {year} {2005}{\natexlab{b}})}\BibitemShut {NoStop}%
\bibitem [{\citenamefont {Schnyder}\ \emph {et~al.}(2008)\citenamefont
  {Schnyder}, \citenamefont {Ryu}, \citenamefont {Furusaki},\ and\
  \citenamefont {Ludwig}}]{schnyder2008classification}%
  \BibitemOpen
  \bibfield  {author} {\bibinfo {author} {\bibfnamefont {A.~P.}\ \bibnamefont
  {Schnyder}}, \bibinfo {author} {\bibfnamefont {S.}~\bibnamefont {Ryu}},
  \bibinfo {author} {\bibfnamefont {A.}~\bibnamefont {Furusaki}},\ and\
  \bibinfo {author} {\bibfnamefont {A.~W.~W.}\ \bibnamefont {Ludwig}},\
  }\bibfield  {title} {\bibinfo {title} {Classification of topological
  insulators and superconductors in three spatial dimensions},\ }\href@noop {}
  {\bibfield  {journal} {\bibinfo  {journal} {Phys. Rev. B}\ }\textbf {\bibinfo
  {volume} {78}},\ \bibinfo {pages} {195125} (\bibinfo {year}
  {2008})}\BibitemShut {NoStop}%
\bibitem [{\citenamefont {Kitaev}(2009)}]{kitaev2009periodic}%
  \BibitemOpen
  \bibfield  {author} {\bibinfo {author} {\bibfnamefont {A.}~\bibnamefont
  {Kitaev}},\ }\bibfield  {title} {\bibinfo {title} {Periodic table for
  topological insulators and superconductors},\ }in\ \href@noop {} {\emph
  {\bibinfo {booktitle} {AIP Conference Proceedings}}},\ Vol.\ \bibinfo
  {volume} {1134}\ (\bibinfo {organization} {AIP},\ \bibinfo {year} {2009})\
  pp.\ \bibinfo {pages} {22--30}\BibitemShut {NoStop}%
\bibitem [{\citenamefont {Ryu}\ \emph {et~al.}(2010)\citenamefont {Ryu},
  \citenamefont {Schnyder}, \citenamefont {Furusaki},\ and\ \citenamefont
  {Ludwig}}]{ryu2010topological}%
  \BibitemOpen
  \bibfield  {author} {\bibinfo {author} {\bibfnamefont {S.}~\bibnamefont
  {Ryu}}, \bibinfo {author} {\bibfnamefont {A.~P.}\ \bibnamefont {Schnyder}},
  \bibinfo {author} {\bibfnamefont {A.}~\bibnamefont {Furusaki}},\ and\
  \bibinfo {author} {\bibfnamefont {A.~W.}\ \bibnamefont {Ludwig}},\ }\bibfield
   {title} {\bibinfo {title} {Topological insulators and superconductors:
  tenfold way and dimensional hierarchy},\ }\href@noop {} {\bibfield  {journal}
  {\bibinfo  {journal} {New J. Phys.}\ }\textbf {\bibinfo {volume} {12}},\
  \bibinfo {pages} {065010} (\bibinfo {year} {2010})}\BibitemShut {NoStop}%
\bibitem [{\citenamefont {Chiu}\ \emph {et~al.}(2016)\citenamefont {Chiu},
  \citenamefont {Teo}, \citenamefont {Schnyder},\ and\ \citenamefont
  {Ryu}}]{chiu2016classification}%
  \BibitemOpen
  \bibfield  {author} {\bibinfo {author} {\bibfnamefont {C.-K.}\ \bibnamefont
  {Chiu}}, \bibinfo {author} {\bibfnamefont {J.~C.~Y.}\ \bibnamefont {Teo}},
  \bibinfo {author} {\bibfnamefont {A.~P.}\ \bibnamefont {Schnyder}},\ and\
  \bibinfo {author} {\bibfnamefont {S.}~\bibnamefont {Ryu}},\ }\bibfield
  {title} {\bibinfo {title} {Classification of topological quantum matter with
  symmetries},\ }\href@noop {} {\bibfield  {journal} {\bibinfo  {journal} {Rev.
  Mod. Phys.}\ }\textbf {\bibinfo {volume} {88}},\ \bibinfo {pages} {035005}
  (\bibinfo {year} {2016})}\BibitemShut {NoStop}%
\bibitem [{\citenamefont {Fu}(2011)}]{fu2011topological}%
  \BibitemOpen
  \bibfield  {author} {\bibinfo {author} {\bibfnamefont {L.}~\bibnamefont
  {Fu}},\ }\bibfield  {title} {\bibinfo {title} {Topological crystalline
  insulators},\ }\href@noop {} {\bibfield  {journal} {\bibinfo  {journal}
  {Phys. Rev. Lett.}\ }\textbf {\bibinfo {volume} {106}},\ \bibinfo {pages}
  {106802} (\bibinfo {year} {2011})}\BibitemShut {NoStop}%
\bibitem [{\citenamefont {Hughes}\ \emph {et~al.}(2011)\citenamefont {Hughes},
  \citenamefont {Prodan},\ and\ \citenamefont
  {Bernevig}}]{hughes2011inversion}%
  \BibitemOpen
  \bibfield  {author} {\bibinfo {author} {\bibfnamefont {T.~L.}\ \bibnamefont
  {Hughes}}, \bibinfo {author} {\bibfnamefont {E.}~\bibnamefont {Prodan}},\
  and\ \bibinfo {author} {\bibfnamefont {B.~A.}\ \bibnamefont {Bernevig}},\
  }\bibfield  {title} {\bibinfo {title} {Inversion-symmetric topological
  insulators},\ }\href@noop {} {\bibfield  {journal} {\bibinfo  {journal}
  {Phys. Rev. B}\ }\textbf {\bibinfo {volume} {83}},\ \bibinfo {pages} {245132}
  (\bibinfo {year} {2011})}\BibitemShut {NoStop}%
\bibitem [{\citenamefont {Hsieh}\ \emph {et~al.}(2012)\citenamefont {Hsieh},
  \citenamefont {Lin}, \citenamefont {Liu}, \citenamefont {Duan}, \citenamefont
  {Bansil},\ and\ \citenamefont {Fu}}]{hsieh2012topological}%
  \BibitemOpen
  \bibfield  {author} {\bibinfo {author} {\bibfnamefont {T.~H.}\ \bibnamefont
  {Hsieh}}, \bibinfo {author} {\bibfnamefont {H.}~\bibnamefont {Lin}}, \bibinfo
  {author} {\bibfnamefont {J.}~\bibnamefont {Liu}}, \bibinfo {author}
  {\bibfnamefont {W.}~\bibnamefont {Duan}}, \bibinfo {author} {\bibfnamefont
  {A.}~\bibnamefont {Bansil}},\ and\ \bibinfo {author} {\bibfnamefont
  {L.}~\bibnamefont {Fu}},\ }\bibfield  {title} {\bibinfo {title} {Topological
  crystalline insulators in the snte material class},\ }\href@noop {}
  {\bibfield  {journal} {\bibinfo  {journal} {Nat. Commun.}\ }\textbf {\bibinfo
  {volume} {3}},\ \bibinfo {pages} {1} (\bibinfo {year} {2012})}\BibitemShut
  {NoStop}%
\bibitem [{\citenamefont {Ando}\ and\ \citenamefont
  {Fu}(2015)}]{ando2015topological}%
  \BibitemOpen
  \bibfield  {author} {\bibinfo {author} {\bibfnamefont {Y.}~\bibnamefont
  {Ando}}\ and\ \bibinfo {author} {\bibfnamefont {L.}~\bibnamefont {Fu}},\
  }\bibfield  {title} {\bibinfo {title} {Topological crystalline insulators and
  topological superconductors: From concepts to materials},\ }\href@noop {}
  {\bibfield  {journal} {\bibinfo  {journal} {Annu. Rev. Condens. Matter
  Phys.}\ }\textbf {\bibinfo {volume} {6}},\ \bibinfo {pages} {361} (\bibinfo
  {year} {2015})}\BibitemShut {NoStop}%
\bibitem [{\citenamefont {Kruthoff}\ \emph {et~al.}(2017)\citenamefont
  {Kruthoff}, \citenamefont {de~Boer}, \citenamefont {van Wezel}, \citenamefont
  {Kane},\ and\ \citenamefont {Slager}}]{kruthoff2017topological}%
  \BibitemOpen
  \bibfield  {author} {\bibinfo {author} {\bibfnamefont {J.}~\bibnamefont
  {Kruthoff}}, \bibinfo {author} {\bibfnamefont {J.}~\bibnamefont {de~Boer}},
  \bibinfo {author} {\bibfnamefont {J.}~\bibnamefont {van Wezel}}, \bibinfo
  {author} {\bibfnamefont {C.~L.}\ \bibnamefont {Kane}},\ and\ \bibinfo
  {author} {\bibfnamefont {R.-J.}\ \bibnamefont {Slager}},\ }\bibfield  {title}
  {\bibinfo {title} {Topological classification of crystalline insulators
  through band structure combinatorics},\ }\href@noop {} {\bibfield  {journal}
  {\bibinfo  {journal} {Phys. Rev. X}\ }\textbf {\bibinfo {volume} {7}},\
  \bibinfo {pages} {041069} (\bibinfo {year} {2017})}\BibitemShut {NoStop}%
\bibitem [{\citenamefont {Po}\ \emph {et~al.}(2017)\citenamefont {Po},
  \citenamefont {Vishwanath},\ and\ \citenamefont {Watanabe}}]{po2017symmetry}%
  \BibitemOpen
  \bibfield  {author} {\bibinfo {author} {\bibfnamefont {H.~C.}\ \bibnamefont
  {Po}}, \bibinfo {author} {\bibfnamefont {A.}~\bibnamefont {Vishwanath}},\
  and\ \bibinfo {author} {\bibfnamefont {H.}~\bibnamefont {Watanabe}},\
  }\bibfield  {title} {\bibinfo {title} {Symmetry-based indicators of band
  topology in the 230 space groups},\ }\href@noop {} {\bibfield  {journal}
  {\bibinfo  {journal} {Nat. Commun.}\ }\textbf {\bibinfo {volume} {8}},\
  \bibinfo {pages} {1} (\bibinfo {year} {2017})}\BibitemShut {NoStop}%
\bibitem [{\citenamefont {Bradlyn}\ \emph {et~al.}(2017)\citenamefont
  {Bradlyn}, \citenamefont {Elcoro}, \citenamefont {Cano}, \citenamefont
  {Vergniory}, \citenamefont {Wang}, \citenamefont {Felser}, \citenamefont
  {Aroyo},\ and\ \citenamefont {Bernevig}}]{bradlyn2017topological}%
  \BibitemOpen
  \bibfield  {author} {\bibinfo {author} {\bibfnamefont {B.}~\bibnamefont
  {Bradlyn}}, \bibinfo {author} {\bibfnamefont {L.}~\bibnamefont {Elcoro}},
  \bibinfo {author} {\bibfnamefont {J.}~\bibnamefont {Cano}}, \bibinfo {author}
  {\bibfnamefont {M.}~\bibnamefont {Vergniory}}, \bibinfo {author}
  {\bibfnamefont {Z.}~\bibnamefont {Wang}}, \bibinfo {author} {\bibfnamefont
  {C.}~\bibnamefont {Felser}}, \bibinfo {author} {\bibfnamefont
  {M.}~\bibnamefont {Aroyo}},\ and\ \bibinfo {author} {\bibfnamefont {B.~A.}\
  \bibnamefont {Bernevig}},\ }\bibfield  {title} {\bibinfo {title} {Topological
  quantum chemistry},\ }\href@noop {} {\bibfield  {journal} {\bibinfo
  {journal} {Nature}\ }\textbf {\bibinfo {volume} {547}},\ \bibinfo {pages}
  {298} (\bibinfo {year} {2017})}\BibitemShut {NoStop}%
\bibitem [{\citenamefont {Schindler}\ \emph {et~al.}(2018)\citenamefont
  {Schindler}, \citenamefont {Cook}, \citenamefont {Vergniory}, \citenamefont
  {Wang}, \citenamefont {Parkin}, \citenamefont {Bernevig},\ and\ \citenamefont
  {Neupert}}]{schindler2018higher}%
  \BibitemOpen
  \bibfield  {author} {\bibinfo {author} {\bibfnamefont {F.}~\bibnamefont
  {Schindler}}, \bibinfo {author} {\bibfnamefont {A.~M.}\ \bibnamefont {Cook}},
  \bibinfo {author} {\bibfnamefont {M.~G.}\ \bibnamefont {Vergniory}}, \bibinfo
  {author} {\bibfnamefont {Z.}~\bibnamefont {Wang}}, \bibinfo {author}
  {\bibfnamefont {S.~S.}\ \bibnamefont {Parkin}}, \bibinfo {author}
  {\bibfnamefont {B.~A.}\ \bibnamefont {Bernevig}},\ and\ \bibinfo {author}
  {\bibfnamefont {T.}~\bibnamefont {Neupert}},\ }\bibfield  {title} {\bibinfo
  {title} {Higher-order topological insulators},\ }\href@noop {} {\bibfield
  {journal} {\bibinfo  {journal} {Sci. Adv.}\ }\textbf {\bibinfo {volume}
  {4}},\ \bibinfo {pages} {eaat0346} (\bibinfo {year} {2018})}\BibitemShut
  {NoStop}%
\bibitem [{\citenamefont {Benalcazar}\ \emph
  {et~al.}(2017{\natexlab{a}})\citenamefont {Benalcazar}, \citenamefont
  {Bernevig},\ and\ \citenamefont {Hughes}}]{benalcazar2017quantized}%
  \BibitemOpen
  \bibfield  {author} {\bibinfo {author} {\bibfnamefont {W.~A.}\ \bibnamefont
  {Benalcazar}}, \bibinfo {author} {\bibfnamefont {B.~A.}\ \bibnamefont
  {Bernevig}},\ and\ \bibinfo {author} {\bibfnamefont {T.~L.}\ \bibnamefont
  {Hughes}},\ }\bibfield  {title} {\bibinfo {title} {Quantized electric
  multipole insulators},\ }\href@noop {} {\bibfield  {journal} {\bibinfo
  {journal} {Science}\ }\textbf {\bibinfo {volume} {357}},\ \bibinfo {pages}
  {61} (\bibinfo {year} {2017}{\natexlab{a}})}\BibitemShut {NoStop}%
\bibitem [{\citenamefont {Benalcazar}\ \emph
  {et~al.}(2017{\natexlab{b}})\citenamefont {Benalcazar}, \citenamefont
  {Bernevig},\ and\ \citenamefont {Hughes}}]{benalcazar2017electric}%
  \BibitemOpen
  \bibfield  {author} {\bibinfo {author} {\bibfnamefont {W.~A.}\ \bibnamefont
  {Benalcazar}}, \bibinfo {author} {\bibfnamefont {B.~A.}\ \bibnamefont
  {Bernevig}},\ and\ \bibinfo {author} {\bibfnamefont {T.~L.}\ \bibnamefont
  {Hughes}},\ }\bibfield  {title} {\bibinfo {title} {Electric multipole
  moments, topological multipole moment pumping, and chiral hinge states in
  crystalline insulators},\ }\href@noop {} {\bibfield  {journal} {\bibinfo
  {journal} {Phys. Rev. B}\ }\textbf {\bibinfo {volume} {96}},\ \bibinfo
  {pages} {245115} (\bibinfo {year} {2017}{\natexlab{b}})}\BibitemShut
  {NoStop}%
\bibitem [{\citenamefont {Khalaf}\ \emph {et~al.}(2019)\citenamefont {Khalaf},
  \citenamefont {Benalcazar}, \citenamefont {Hughes},\ and\ \citenamefont
  {Queiroz}}]{khalaf2019boundary}%
  \BibitemOpen
  \bibfield  {author} {\bibinfo {author} {\bibfnamefont {E.}~\bibnamefont
  {Khalaf}}, \bibinfo {author} {\bibfnamefont {W.~A.}\ \bibnamefont
  {Benalcazar}}, \bibinfo {author} {\bibfnamefont {T.~L.}\ \bibnamefont
  {Hughes}},\ and\ \bibinfo {author} {\bibfnamefont {R.}~\bibnamefont
  {Queiroz}},\ }\bibfield  {title} {\bibinfo {title} {Boundary-obstructed
  topological phases},\ }\href@noop {} {\bibfield  {journal} {\bibinfo
  {journal} {arXiv preprint arXiv:1908.00011}\ } (\bibinfo {year}
  {2019})}\BibitemShut {NoStop}%
\bibitem [{\citenamefont {Prodan}\ \emph {et~al.}(2010)\citenamefont {Prodan},
  \citenamefont {Hughes},\ and\ \citenamefont {Bernevig}}]{prodan2010}%
  \BibitemOpen
  \bibfield  {author} {\bibinfo {author} {\bibfnamefont {E.}~\bibnamefont
  {Prodan}}, \bibinfo {author} {\bibfnamefont {T.~L.}\ \bibnamefont {Hughes}},\
  and\ \bibinfo {author} {\bibfnamefont {B.~A.}\ \bibnamefont {Bernevig}},\
  }\bibfield  {title} {\bibinfo {title} {Entanglement spectrum of a disordered
  topological chern insulator},\ }\href@noop {} {\bibfield  {journal} {\bibinfo
   {journal} {Phys. Rev. Lett.}\ }\textbf {\bibinfo {volume} {105}},\ \bibinfo
  {pages} {115501} (\bibinfo {year} {2010})}\BibitemShut {NoStop}%
\bibitem [{\citenamefont {Prodan}\ and\ \citenamefont
  {Schulz-Baldes}(2016{\natexlab{a}})}]{prodan2016non}%
  \BibitemOpen
  \bibfield  {author} {\bibinfo {author} {\bibfnamefont {E.}~\bibnamefont
  {Prodan}}\ and\ \bibinfo {author} {\bibfnamefont {H.}~\bibnamefont
  {Schulz-Baldes}},\ }\bibfield  {title} {\bibinfo {title} {Non-commutative odd
  chern numbers and topological phases of disordered chiral systems},\
  }\href@noop {} {\bibfield  {journal} {\bibinfo  {journal} {J. Funct. Anal.}\
  }\textbf {\bibinfo {volume} {271}},\ \bibinfo {pages} {1150} (\bibinfo {year}
  {2016}{\natexlab{a}})}\BibitemShut {NoStop}%
\bibitem [{\citenamefont {Prodan}\ and\ \citenamefont
  {Schulz-Baldes}(2016{\natexlab{b}})}]{prodan2016bulk}%
  \BibitemOpen
  \bibfield  {author} {\bibinfo {author} {\bibfnamefont {E.}~\bibnamefont
  {Prodan}}\ and\ \bibinfo {author} {\bibfnamefont {H.}~\bibnamefont
  {Schulz-Baldes}},\ }\href@noop {} {\emph {\bibinfo {title} {Bulk and boundary
  invariants for complex topological insulators}}},\ Mathematical Physics
  Studies\ (\bibinfo  {publisher} {Springer},\ \bibinfo {year}
  {2016})\BibitemShut {NoStop}%
\bibitem [{\citenamefont {Prodan}(2017)}]{prodan2017computational}%
  \BibitemOpen
  \bibfield  {author} {\bibinfo {author} {\bibfnamefont {E.}~\bibnamefont
  {Prodan}},\ }\href@noop {} {\emph {\bibinfo {title} {A computational
  non-commutative geometry program for disordered topological insulators}}},\
  \bibinfo {series} {SpringerBriefs in Mathematical Physics}, Vol.~\bibinfo
  {volume} {23}\ (\bibinfo  {publisher} {Springer},\ \bibinfo {year}
  {2017})\BibitemShut {NoStop}%
\bibitem [{\citenamefont {Prodan}\ \emph {et~al.}(2013)\citenamefont {Prodan},
  \citenamefont {Leung},\ and\ \citenamefont {Bellissard}}]{prodan2013non}%
  \BibitemOpen
  \bibfield  {author} {\bibinfo {author} {\bibfnamefont {E.}~\bibnamefont
  {Prodan}}, \bibinfo {author} {\bibfnamefont {B.}~\bibnamefont {Leung}},\ and\
  \bibinfo {author} {\bibfnamefont {J.}~\bibnamefont {Bellissard}},\ }\bibfield
   {title} {\bibinfo {title} {The non-commutative nth-chern number ($n\geq
  1$)},\ }\href@noop {} {\bibfield  {journal} {\bibinfo  {journal} {J. Phys.
  A}\ }\textbf {\bibinfo {volume} {46}},\ \bibinfo {pages} {485202} (\bibinfo
  {year} {2013})}\BibitemShut {NoStop}%
\bibitem [{\citenamefont {Mondragon-Shem}\ \emph {et~al.}(2014)\citenamefont
  {Mondragon-Shem}, \citenamefont {Hughes}, \citenamefont {Song},\ and\
  \citenamefont {Prodan}}]{mondragon2014topological}%
  \BibitemOpen
  \bibfield  {author} {\bibinfo {author} {\bibfnamefont {I.}~\bibnamefont
  {Mondragon-Shem}}, \bibinfo {author} {\bibfnamefont {T.~L.}\ \bibnamefont
  {Hughes}}, \bibinfo {author} {\bibfnamefont {J.}~\bibnamefont {Song}},\ and\
  \bibinfo {author} {\bibfnamefont {E.}~\bibnamefont {Prodan}},\ }\bibfield
  {title} {\bibinfo {title} {Topological criticality in the chiral-symmetric
  {AIII} class at strong disorder},\ }\href@noop {} {\bibfield  {journal}
  {\bibinfo  {journal} {Phys. Rev. Lett.}\ }\textbf {\bibinfo {volume} {113}},\
  \bibinfo {pages} {046802} (\bibinfo {year} {2014})}\BibitemShut {NoStop}%
\bibitem [{\citenamefont {Loring}\ and\ \citenamefont
  {Hastings}(2011)}]{loring2011}%
  \BibitemOpen
  \bibfield  {author} {\bibinfo {author} {\bibfnamefont {T.~A.}\ \bibnamefont
  {Loring}}\ and\ \bibinfo {author} {\bibfnamefont {M.~B.}\ \bibnamefont
  {Hastings}},\ }\bibfield  {title} {\bibinfo {title} {Disordered topological
  insulators via {C}*-algebras},\ }\href@noop {} {\bibfield  {journal}
  {\bibinfo  {journal} {EPL}\ }\textbf {\bibinfo {volume} {92}},\ \bibinfo
  {pages} {67004} (\bibinfo {year} {2011})}\BibitemShut {NoStop}%
\bibitem [{\citenamefont {Schulz-Baldes}(2016)}]{schulz2016topological}%
  \BibitemOpen
  \bibfield  {author} {\bibinfo {author} {\bibfnamefont {H.}~\bibnamefont
  {Schulz-Baldes}},\ }\bibfield  {title} {\bibinfo {title} {H},\ }\href@noop {}
  {\bibfield  {journal} {\bibinfo  {journal} {Jahresber. Dtsch. Math.-Ver.}\
  }\textbf {\bibinfo {volume} {118}},\ \bibinfo {pages} {247} (\bibinfo {year}
  {2016})}\BibitemShut {NoStop}%
\bibitem [{\citenamefont {Song}\ and\ \citenamefont
  {Prodan}(2015)}]{song2015quantization}%
  \BibitemOpen
  \bibfield  {author} {\bibinfo {author} {\bibfnamefont {J.}~\bibnamefont
  {Song}}\ and\ \bibinfo {author} {\bibfnamefont {E.}~\bibnamefont {Prodan}},\
  }\bibfield  {title} {\bibinfo {title} {Quantization of topological invariants
  under symmetry-breaking disorder},\ }\href@noop {} {\bibfield  {journal}
  {\bibinfo  {journal} {Phys. Rev. B}\ }\textbf {\bibinfo {volume} {92}},\
  \bibinfo {pages} {195119} (\bibinfo {year} {2015})}\BibitemShut {NoStop}%
\bibitem [{\citenamefont {Diez}\ \emph {et~al.}(2015)\citenamefont {Diez},
  \citenamefont {Pikulin}, \citenamefont {Fulga},\ and\ \citenamefont
  {Tworzyd{\l}o}}]{diez2015extended}%
  \BibitemOpen
  \bibfield  {author} {\bibinfo {author} {\bibfnamefont {M.}~\bibnamefont
  {Diez}}, \bibinfo {author} {\bibfnamefont {D.~I.}\ \bibnamefont {Pikulin}},
  \bibinfo {author} {\bibfnamefont {I.~C.}\ \bibnamefont {Fulga}},\ and\
  \bibinfo {author} {\bibfnamefont {J.}~\bibnamefont {Tworzyd{\l}o}},\
  }\bibfield  {title} {\bibinfo {title} {Extended topological group structure
  due to average reflection symmetry},\ }\href@noop {} {\bibfield  {journal}
  {\bibinfo  {journal} {New J. Phys.}\ }\textbf {\bibinfo {volume} {17}},\
  \bibinfo {pages} {043014} (\bibinfo {year} {2015})}\BibitemShut {NoStop}%
\bibitem [{\citenamefont {Mondragon-Shem}\ and\ \citenamefont
  {Hughes}(2019)}]{mondragon2019robust}%
  \BibitemOpen
  \bibfield  {author} {\bibinfo {author} {\bibfnamefont {I.}~\bibnamefont
  {Mondragon-Shem}}\ and\ \bibinfo {author} {\bibfnamefont {T.~L.}\
  \bibnamefont {Hughes}},\ }\bibfield  {title} {\bibinfo {title} {Robust
  topological invariants of topological crystalline phases in the presence of
  impurities},\ }\href@noop {} {\bibfield  {journal} {\bibinfo  {journal}
  {arXiv preprint arXiv:1906.11847}\ } (\bibinfo {year} {2019})}\BibitemShut
  {NoStop}%
\bibitem [{\citenamefont {Fu}\ and\ \citenamefont
  {Kane}(2012)}]{fu2012topology}%
  \BibitemOpen
  \bibfield  {author} {\bibinfo {author} {\bibfnamefont {L.}~\bibnamefont
  {Fu}}\ and\ \bibinfo {author} {\bibfnamefont {C.~L.}\ \bibnamefont {Kane}},\
  }\bibfield  {title} {\bibinfo {title} {Topology, delocalization via average
  symmetry and the symplectic anderson transition},\ }\href@noop {} {\bibfield
  {journal} {\bibinfo  {journal} {Phys. Rev. Lett.}\ }\textbf {\bibinfo
  {volume} {109}},\ \bibinfo {pages} {246605} (\bibinfo {year}
  {2012})}\BibitemShut {NoStop}%
\bibitem [{\citenamefont {Fulga}\ \emph {et~al.}(2014)\citenamefont {Fulga},
  \citenamefont {van Heck}, \citenamefont {Edge},\ and\ \citenamefont
  {Akhmerov}}]{fulga2014statistical}%
  \BibitemOpen
  \bibfield  {author} {\bibinfo {author} {\bibfnamefont {I.~C.}\ \bibnamefont
  {Fulga}}, \bibinfo {author} {\bibfnamefont {B.}~\bibnamefont {van Heck}},
  \bibinfo {author} {\bibfnamefont {J.~M.}\ \bibnamefont {Edge}},\ and\
  \bibinfo {author} {\bibfnamefont {A.~R.}\ \bibnamefont {Akhmerov}},\
  }\bibfield  {title} {\bibinfo {title} {Statistical topological insulators},\
  }\href@noop {} {\bibfield  {journal} {\bibinfo  {journal} {Phys. Rev. B}\
  }\textbf {\bibinfo {volume} {89}},\ \bibinfo {pages} {155424} (\bibinfo
  {year} {2014})}\BibitemShut {NoStop}%
\bibitem [{\citenamefont {Serra-Garcia}\ \emph {et~al.}(2018)\citenamefont
  {Serra-Garcia}, \citenamefont {Peri}, \citenamefont {S{\"u}sstrunk},
  \citenamefont {Bilal}, \citenamefont {Larsen}, \citenamefont {Villanueva},\
  and\ \citenamefont {Huber}}]{serra2018}%
  \BibitemOpen
  \bibfield  {author} {\bibinfo {author} {\bibfnamefont {M.}~\bibnamefont
  {Serra-Garcia}}, \bibinfo {author} {\bibfnamefont {V.}~\bibnamefont {Peri}},
  \bibinfo {author} {\bibfnamefont {R.}~\bibnamefont {S{\"u}sstrunk}}, \bibinfo
  {author} {\bibfnamefont {O.~R.}\ \bibnamefont {Bilal}}, \bibinfo {author}
  {\bibfnamefont {T.}~\bibnamefont {Larsen}}, \bibinfo {author} {\bibfnamefont
  {L.~G.}\ \bibnamefont {Villanueva}},\ and\ \bibinfo {author} {\bibfnamefont
  {S.~D.}\ \bibnamefont {Huber}},\ }\bibfield  {title} {\bibinfo {title}
  {Observation of a phononic quadrupole topological insulator},\ }\href@noop {}
  {\bibfield  {journal} {\bibinfo  {journal} {Nature}\ }\textbf {\bibinfo
  {volume} {555}},\ \bibinfo {pages} {342} (\bibinfo {year}
  {2018})}\BibitemShut {NoStop}%
\bibitem [{\citenamefont {Peterson}\ \emph {et~al.}(2018)\citenamefont
  {Peterson}, \citenamefont {Benalcazar}, \citenamefont {Hughes},\ and\
  \citenamefont {Bahl}}]{peterson2018}%
  \BibitemOpen
  \bibfield  {author} {\bibinfo {author} {\bibfnamefont {C.~W.}\ \bibnamefont
  {Peterson}}, \bibinfo {author} {\bibfnamefont {W.~A.}\ \bibnamefont
  {Benalcazar}}, \bibinfo {author} {\bibfnamefont {T.~L.}\ \bibnamefont
  {Hughes}},\ and\ \bibinfo {author} {\bibfnamefont {G.}~\bibnamefont {Bahl}},\
  }\bibfield  {title} {\bibinfo {title} {A quantized microwave quadrupole
  insulator with topologically protected corner states},\ }\href@noop {}
  {\bibfield  {journal} {\bibinfo  {journal} {Nature}\ }\textbf {\bibinfo
  {volume} {555}},\ \bibinfo {pages} {346} (\bibinfo {year}
  {2018})}\BibitemShut {NoStop}%
\bibitem [{\citenamefont {Imhof}\ \emph {et~al.}(2018)\citenamefont {Imhof},
  \citenamefont {Berger}, \citenamefont {Bayer}, \citenamefont {Brehm},
  \citenamefont {Molenkamp}, \citenamefont {Kiessling}, \citenamefont
  {Schindler}, \citenamefont {Lee}, \citenamefont {Greiter}, \citenamefont
  {Neupert} \emph {et~al.}}]{imhof2018}%
  \BibitemOpen
  \bibfield  {author} {\bibinfo {author} {\bibfnamefont {S.}~\bibnamefont
  {Imhof}}, \bibinfo {author} {\bibfnamefont {C.}~\bibnamefont {Berger}},
  \bibinfo {author} {\bibfnamefont {F.}~\bibnamefont {Bayer}}, \bibinfo
  {author} {\bibfnamefont {J.}~\bibnamefont {Brehm}}, \bibinfo {author}
  {\bibfnamefont {L.~W.}\ \bibnamefont {Molenkamp}}, \bibinfo {author}
  {\bibfnamefont {T.}~\bibnamefont {Kiessling}}, \bibinfo {author}
  {\bibfnamefont {F.}~\bibnamefont {Schindler}}, \bibinfo {author}
  {\bibfnamefont {C.~H.}\ \bibnamefont {Lee}}, \bibinfo {author} {\bibfnamefont
  {M.}~\bibnamefont {Greiter}}, \bibinfo {author} {\bibfnamefont
  {T.}~\bibnamefont {Neupert}}, \emph {et~al.},\ }\bibfield  {title} {\bibinfo
  {title} {Topolectrical-circuit realization of topological corner modes},\
  }\href@noop {} {\bibfield  {journal} {\bibinfo  {journal} {Nature Physics}\
  }\textbf {\bibinfo {volume} {14}},\ \bibinfo {pages} {925} (\bibinfo {year}
  {2018})}\BibitemShut {NoStop}%
\bibitem [{\citenamefont {Meier}\ \emph {et~al.}(2018)\citenamefont {Meier},
  \citenamefont {An}, \citenamefont {Dauphin}, \citenamefont {Maffei},
  \citenamefont {Massignan}, \citenamefont {Hughes},\ and\ \citenamefont
  {Gadway}}]{meier2018}%
  \BibitemOpen
  \bibfield  {author} {\bibinfo {author} {\bibfnamefont {E.~J.}\ \bibnamefont
  {Meier}}, \bibinfo {author} {\bibfnamefont {F.~A.}\ \bibnamefont {An}},
  \bibinfo {author} {\bibfnamefont {A.}~\bibnamefont {Dauphin}}, \bibinfo
  {author} {\bibfnamefont {M.}~\bibnamefont {Maffei}}, \bibinfo {author}
  {\bibfnamefont {P.}~\bibnamefont {Massignan}}, \bibinfo {author}
  {\bibfnamefont {T.~L.}\ \bibnamefont {Hughes}},\ and\ \bibinfo {author}
  {\bibfnamefont {B.}~\bibnamefont {Gadway}},\ }\bibfield  {title} {\bibinfo
  {title} {Observation of the topological anderson insulator in disordered
  atomic wires},\ }\href@noop {} {\bibfield  {journal} {\bibinfo  {journal}
  {Science}\ }\textbf {\bibinfo {volume} {362}},\ \bibinfo {pages} {929}
  (\bibinfo {year} {2018})}\BibitemShut {NoStop}%
\bibitem [{\citenamefont {Alexandradinata}\ \emph {et~al.}(2014)\citenamefont
  {Alexandradinata}, \citenamefont {Dai},\ and\ \citenamefont
  {Bernevig}}]{alexandradinata2014wilson}%
  \BibitemOpen
  \bibfield  {author} {\bibinfo {author} {\bibfnamefont {A.}~\bibnamefont
  {Alexandradinata}}, \bibinfo {author} {\bibfnamefont {X.}~\bibnamefont
  {Dai}},\ and\ \bibinfo {author} {\bibfnamefont {B.~A.}\ \bibnamefont
  {Bernevig}},\ }\bibfield  {title} {\bibinfo {title} {Wilson-loop
  characterization of inversion-symmetric topological insulators},\ }\href@noop
  {} {\bibfield  {journal} {\bibinfo  {journal} {Phys. Rev. B}\ }\textbf
  {\bibinfo {volume} {89}},\ \bibinfo {pages} {155114} (\bibinfo {year}
  {2014})}\BibitemShut {NoStop}%
\bibitem [{\citenamefont {Fidkowski}\ \emph {et~al.}(2011)\citenamefont
  {Fidkowski}, \citenamefont {Jackson},\ and\ \citenamefont
  {Klich}}]{fidkowski2011model}%
  \BibitemOpen
  \bibfield  {author} {\bibinfo {author} {\bibfnamefont {L.}~\bibnamefont
  {Fidkowski}}, \bibinfo {author} {\bibfnamefont {T.~S.}\ \bibnamefont
  {Jackson}},\ and\ \bibinfo {author} {\bibfnamefont {I.}~\bibnamefont
  {Klich}},\ }\bibfield  {title} {\bibinfo {title} {Model characterization of
  gapless edge modes of topological insulators using intermediate
  brillouin-zone functions},\ }\href@noop {} {\bibfield  {journal} {\bibinfo
  {journal} {Phys. Rev. Lett.}\ }\textbf {\bibinfo {volume} {107}},\ \bibinfo
  {pages} {036601} (\bibinfo {year} {2011})}\BibitemShut {NoStop}%
\bibitem [{\citenamefont {Su}\ \emph {et~al.}(1979)\citenamefont {Su},
  \citenamefont {Schrieffer},\ and\ \citenamefont {Heeger}}]{su1979solitons}%
  \BibitemOpen
  \bibfield  {author} {\bibinfo {author} {\bibfnamefont {W.~P.}\ \bibnamefont
  {Su}}, \bibinfo {author} {\bibfnamefont {J.~R.}\ \bibnamefont {Schrieffer}},\
  and\ \bibinfo {author} {\bibfnamefont {A.~J.}\ \bibnamefont {Heeger}},\
  }\bibfield  {title} {\bibinfo {title} {Solitons in polyacetylene},\
  }\href@noop {} {\bibfield  {journal} {\bibinfo  {journal} {Phys. Rev. Lett.}\
  }\textbf {\bibinfo {volume} {42}},\ \bibinfo {pages} {1698} (\bibinfo {year}
  {1979})}\BibitemShut {NoStop}%
\bibitem [{\citenamefont {Schulz-Baldes}\ and\ \citenamefont
  {Teufel}(2013)}]{schulz2013orbital}%
  \BibitemOpen
  \bibfield  {author} {\bibinfo {author} {\bibfnamefont {H.}~\bibnamefont
  {Schulz-Baldes}}\ and\ \bibinfo {author} {\bibfnamefont {S.}~\bibnamefont
  {Teufel}},\ }\bibfield  {title} {\bibinfo {title} {Orbital polarization and
  magnetization for independent particles in disordered media},\ }\href@noop {}
  {\bibfield  {journal} {\bibinfo  {journal} {Commun. Math. Phys.}\ }\textbf
  {\bibinfo {volume} {319}},\ \bibinfo {pages} {649} (\bibinfo {year}
  {2013})}\BibitemShut {NoStop}%
\bibitem [{\citenamefont {Resta}(1998)}]{resta1998quantum}%
  \BibitemOpen
  \bibfield  {author} {\bibinfo {author} {\bibfnamefont {R.}~\bibnamefont
  {Resta}},\ }\bibfield  {title} {\bibinfo {title} {Quantum-mechanical position
  operator in extended systems},\ }\href@noop {} {\bibfield  {journal}
  {\bibinfo  {journal} {Phys. Rev. Lett.}\ }\textbf {\bibinfo {volume} {80}},\
  \bibinfo {pages} {1800} (\bibinfo {year} {1998})}\BibitemShut {NoStop}%
\bibitem [{\citenamefont {Resta}\ and\ \citenamefont
  {Sorella}(1999)}]{resta1999electron}%
  \BibitemOpen
  \bibfield  {author} {\bibinfo {author} {\bibfnamefont {R.}~\bibnamefont
  {Resta}}\ and\ \bibinfo {author} {\bibfnamefont {S.}~\bibnamefont
  {Sorella}},\ }\bibfield  {title} {\bibinfo {title} {Electron localization in
  the insulating state},\ }\href@noop {} {\bibfield  {journal} {\bibinfo
  {journal} {Physical Review Letters}\ }\textbf {\bibinfo {volume} {82}},\
  \bibinfo {pages} {370} (\bibinfo {year} {1999})}\BibitemShut {NoStop}%
\bibitem [{\citenamefont {Fisher}(1994)}]{fisher1994random}%
  \BibitemOpen
  \bibfield  {author} {\bibinfo {author} {\bibfnamefont {D.~S.}\ \bibnamefont
  {Fisher}},\ }\bibfield  {title} {\bibinfo {title} {Random antiferromagnetic
  quantum spin chains},\ }\href@noop {} {\bibfield  {journal} {\bibinfo
  {journal} {Phys. Rev. B}\ }\textbf {\bibinfo {volume} {50}},\ \bibinfo
  {pages} {3799} (\bibinfo {year} {1994})}\BibitemShut {NoStop}%
\bibitem [{Note1()}]{Note1}%
  \BibitemOpen
  \bibinfo {note} {See Supplemental Material at [URL will be inserted by
  publisher] for a detailed description of the RG procedure and proof that the
  RG preserves the structure of the $\pi $-flux ladder}\BibitemShut {NoStop}%
\bibitem [{\citenamefont {Mila}\ and\ \citenamefont
  {Schmidt}(2011)}]{mila2011strong}%
  \BibitemOpen
  \bibfield  {author} {\bibinfo {author} {\bibfnamefont {F.}~\bibnamefont
  {Mila}}\ and\ \bibinfo {author} {\bibfnamefont {K.~P.}\ \bibnamefont
  {Schmidt}},\ }\bibfield  {title} {\bibinfo {title} {Strong-coupling expansion
  and effective hamiltonians},\ }in\ \href@noop {} {\emph {\bibinfo {booktitle}
  {Introduction to Frustrated Magnetism}}}\ (\bibinfo  {publisher} {Springer},\
  \bibinfo {year} {2011})\ pp.\ \bibinfo {pages} {537--559}\BibitemShut
  {NoStop}%
\end{thebibliography}%
